\documentclass[aps,prl,twocolumn,showpacs,superscriptaddress,floatfix,nofootinbib]{revtex4-1}
\usepackage{graphicx}
\usepackage{amsmath}
\usepackage{epsfig}
\usepackage{helvet}
\usepackage{amssymb}

\def\L{\mathcal{L}}
\def\S{\mathcal{S}}

\def\HI{H_{\mbox{\tiny Ising}}}
\def\HXX{H_{\mbox{\tiny XX}}}

\def\tGamma{\Gamma^{\triangleleft}}
\def\to{o^{\triangleleft}}
\def\tphi{\phi^{\triangleleft}}

\begin{document}

\title{Non-local scaling operators with entanglement renormalization}

\author{G. Evenbly}
\affiliation{School of Mathematics and Physics, the University of Queensland, QLD 4072, Australia}

\author{P. Corboz}
\affiliation{School of Mathematics and Physics, the University of Queensland, QLD 4072, Australia}

\author{G. Vidal}
\affiliation{School of Mathematics and Physics, the University of Queensland, QLD 4072, Australia}
\date{\today}

\begin{abstract}
The multi-scale entanglement renormalization ansatz (MERA) can be used, in its scale invariant version, to describe the ground state of a lattice system at a quantum critical point. From the scale invariant MERA one can determine the local scaling operators of the model. Here we show that, in the presence of a global symmetry $\mathcal{G}$, it is also possible to determine a class of non-local scaling operators. Each operator consists, for a given group element $g\in\mathcal{G}$, of a semi-infinite string $\tGamma_g$ with a local operator $\varphi$ attached to its open end. In the case of the quantum Ising model, $\mathcal{G}= \mathbb{Z}_2$, they correspond to the disorder operator $\mu$, the fermionic operators $\psi$ and $\bar{\psi}$, and all their descendants. 
Together with the local scaling operators identity $\mathbb{I}$, spin $\sigma$ and energy $\epsilon$, the fermionic and disorder scaling operators $\psi$, $\bar{\psi}$ and $\mu$ are the complete list of primary fields of the Ising CFT. Thefore the scale invariant MERA allows us to characterize all the conformal towers of this CFT. 
\end{abstract}

\pacs{03.67.--a, 05.50.+q, 11.25.Hf}

\maketitle

The multi-scale entanglement renormalization ansatz (MERA) \cite{ER,MERA} is a tensor network introduced to efficiently represent ground states and low energy subspaces of quantum many-body systems on a lattice. It is based on a real-space renormalization group (RG) technique known as \emph{entanglement renormalization} \cite{ER}, that employs unitary tensors (\emph{disentanglers}) to remove short-range entanglement from the system at each RG iteration. The removal of entanglement is a key difference with other real-space RG techniques, such as Wilson's ground breaking numerical RG (NRG) for the Kondo problem \cite{Wilson} or White's extremely succesful density matrix RG (DMRG) \cite{DMRG} for arbitrary one-dimensional systems. At a fixed point of the RG flow, it produces a representation that is explicitly scale invariant: the \emph{scale invariant} MERA. This ansatz is characterized by only a small number of tensors and can be used to describe systems with topological order (at a non-critical RG fixed point) \cite{Topo} as well as continuous quantum phase transitions (corresponding to a critical RG fixed point) \cite{ER,MERA,Free,Transfer,MERACFT,Fazio,Boundary}.

Here we shall be concerned with the characterization of a (scale invariant) quantum critical point with the scale invariant MERA \cite{ER, MERA, Free, Transfer, MERACFT, Fazio, Boundary}. Evidence that the scale-invariant MERA is capable of describing critical ground states was first presented in Ref. \cite{ER} for the quantum Ising model and in Ref. \cite{Free} for non-interacting systems of fermions and bosons. On the other hand it was argued that this ansatz naturally reproduces two important aspects of critical ground states: the logarithmic scaling for the entanglement entropy of a block of contiguous sites (in one-dimensional systems) and the power-law decay of correlations \cite{MERA}. The latter was seen to follow from the fact that a two-point correlator $C(s_1,s_2)$ between two points separated a distance $r=|s_1-s_2|$ is obtained after $O(\log (r))$ applications of a fixed superoperator that introduces a constant factor $z<1$ after each application, and therefore 
\begin{equation}
	C_2(s_1,s_2) \approx z^{\log (r)} = r^{-q},~~~~q \equiv - \log z.
\end{equation}
This result was formalized in Ref. \cite{Transfer} by relating the possible values of the factor $z$ with the eigenvalues of that superoperator, and by identifying the \emph{scaling operators} of the theory with the corresponding eigenoperators. A connection between the scale invariant MERA and the conformal field theory (CFT) underlying a quantum critical point was then established in Ref. \cite{MERACFT}, including a way to extract the conformal data: central charge, primary fields, and their scaling dimensions and operator product expansion (OPE) \cite{CFT}. However, the analysis of Refs. \cite{Transfer,MERACFT} was only concerned with \emph{local} scaling operators. For instance, for the quantum Ising model, Ref. \cite{MERACFT} identified the primary fields identity $\mathbb{I}$, energy $\epsilon$ and spin $\sigma$, as well as some of their descendants, all of which were expressed as an operator acting on two contiguous sites of a coarse-grained lattice. Instead, \emph{non-local} scaling operators were not considered. One reason is that entanglement renormalization, being based on locally coarse-graining the system, has a computational cost that grows exponentially with the size of the support of the operators under consideration. 

In this paper we show that the scale invariant MERA can be used to characterize a whole class of non-local scaling operators of a critical lattice model. We consider a quantum spin chain whose Hamiltonian $H$ is invariant under a symmetry group $\mathcal{G}$,
\begin{equation}
 \Gamma_g ~H~ \Gamma_g^{\dagger} = H, ~~~~\forall g \in \mathcal{G},
 \label{eq:symm}
\end{equation}
where $\Gamma_g \equiv \cdots V_g \otimes V_g \otimes V_g \cdots$ is an infinite string of copies of a matrix $V_g$, with $V_g$ a unitary representation of $\mathcal{G}$. We shall see that, by incorporating the symmetry $\mathcal{G}$ into the MERA, it is possible to study non-local operators that have a semi-infinite string of $V_g$'s. 
Important examples of such non-local operators are the disorder operator $\mu$ and the fermionic operators $\psi$ and $\bar{\psi}$ of the quantum Ising model, which are associated to the low energy spectrum of a chain with anti-periodic boundary conditions.

\begin{figure}[!htbp]
\begin{center}
\includegraphics[width=8cm]{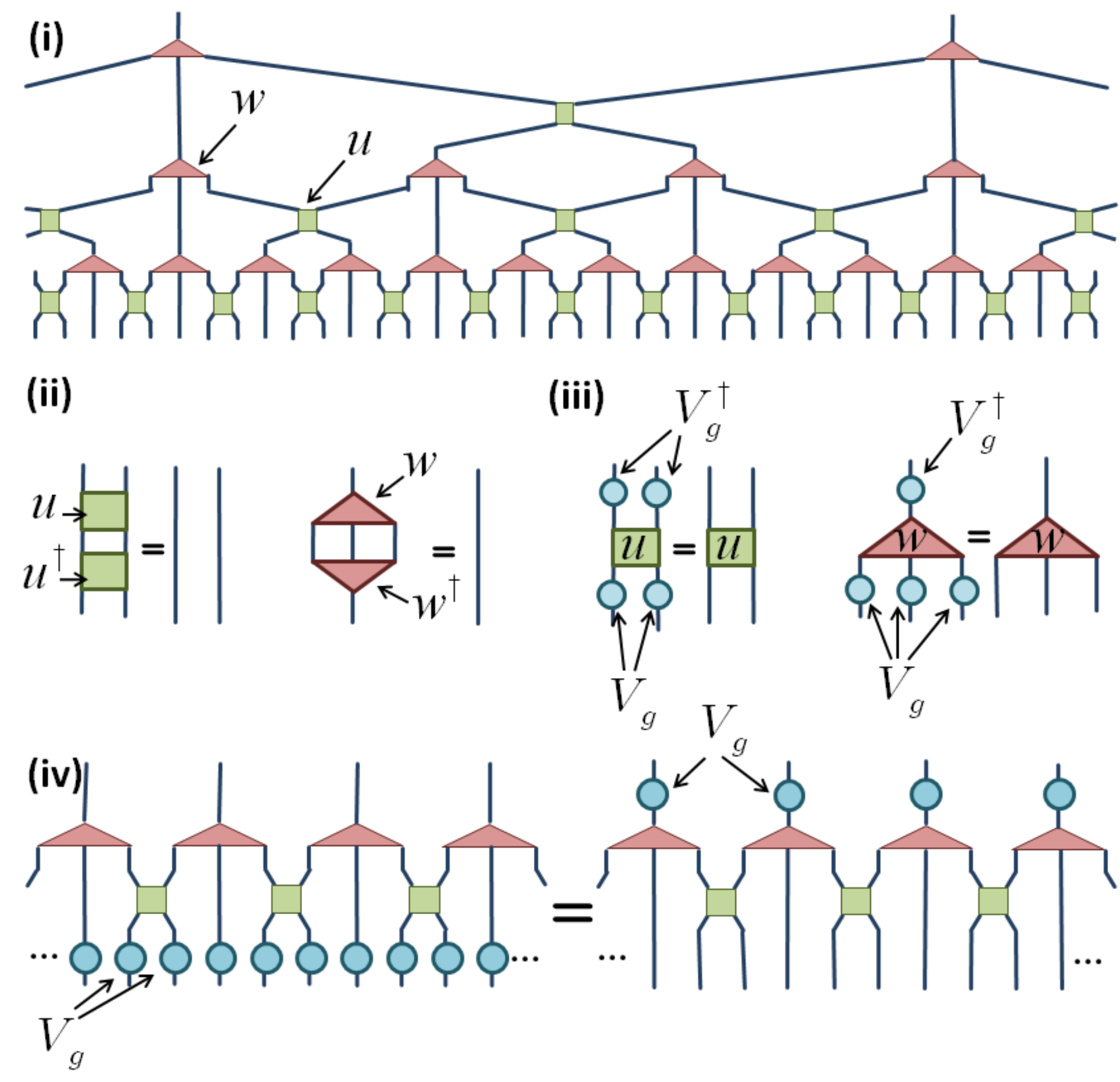}
\caption{ (i) Scale invariant MERA, characterized by a disentangler $u$ and an isometry $w$ that are copied throughout the ansatz. (ii) These tensors are isometric, meaning that $u^{\dagger}u = I$, $w^{\dagger}w = I$. (iii) We choose $u$ and $w$ to be symmetric, Eq. \ref{eq:symmetric}. (iv) As a result, the infinite string $\tGamma_g$ commutes with a layer of disentanglers and isometries. In other words, $\tGamma_g$ is invariant under coarse-graining.}
\label{fig:symMERA}
\end{center}
\end{figure}

\begin{figure}[!htbp]
\begin{center}
\includegraphics[width=8cm]{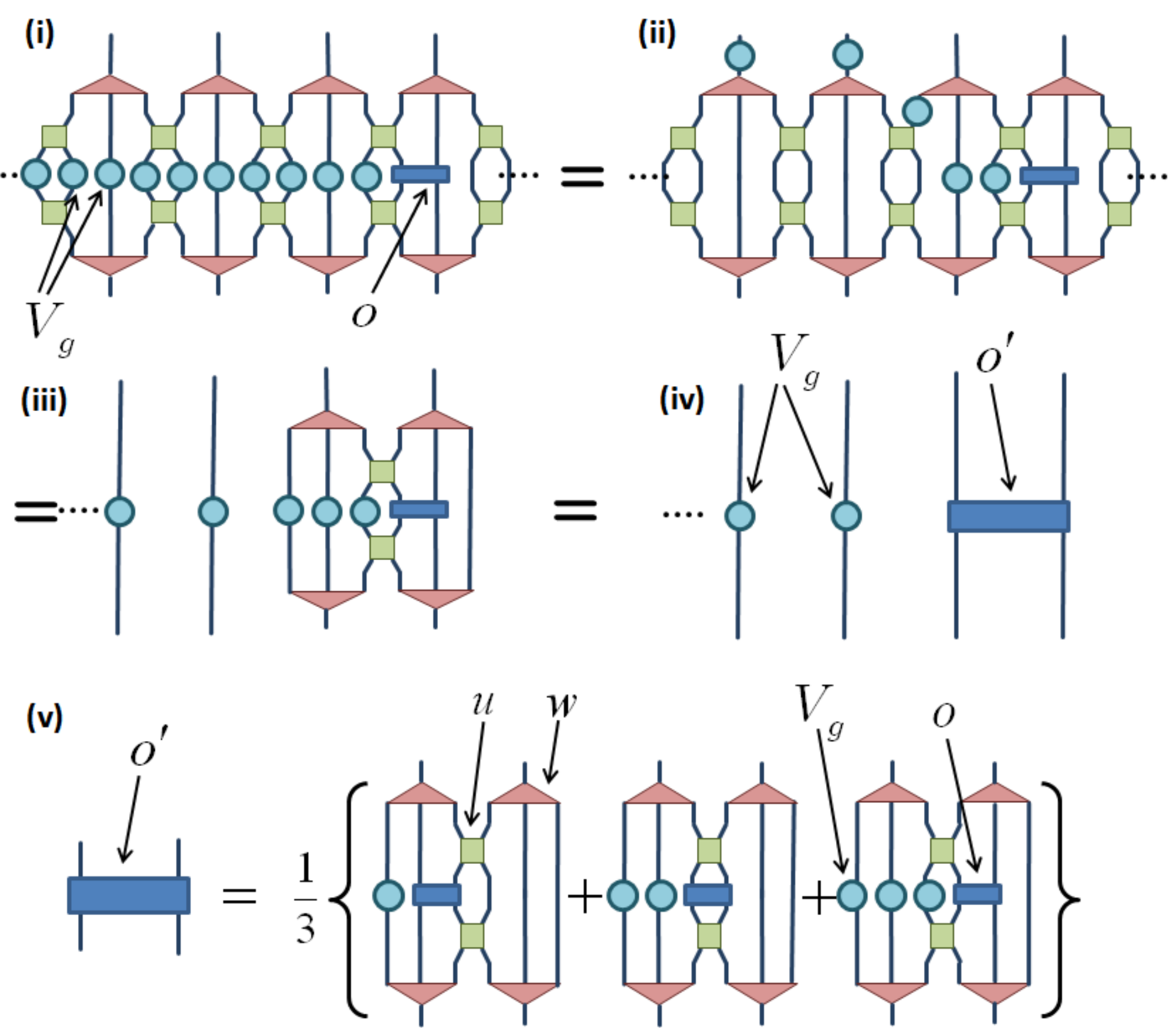}
\caption{(i) Coarse-graining of a non-local operator $\to = \tGamma_g\otimes o$. (ii) Most of the string $\tGamma_g$ of $V_g$'s commutes with the coarse-graining thanks to Eq. \ref{eq:symmetric}. (iii) Then we can remove most of disentanglers and isometries using Fig. \ref{fig:symMERA}(ii). (iv) $o'$ is defined in term of $o$, $u$, $w$ and $V_g$. (v) Scaling superoperator $\mathcal{S}_g$, $o'=S_g(o)$, for the local part of a non-local operator $\to_g$ in Eq. \ref{eq:tog}, see Eqs. \ref{eq:Stog}-\ref{eq:Stog3}. Notice the average over the three possible ways in which $o$ can be coarse-grained. In the case of $g=\mathbb{I}$, we have $V_g = \mathbb{I}$, so that that $\to_{\mathbb{I}}$ is simply a local operator and we recover the `usual' scaling superoperator $\mathcal{S}$ of Eqs. \ref{eq:S}-\ref{eq:SS} (see Fig. 1 of Ref. \cite{MERACFT} for further details).}
\label{fig:CoarseGrain}
\end{center}
\end{figure}

\emph{Local scaling operators.}--- Recall that the scale invariant MERA is made of copies of a unique pair of bulk tensors, namely a disentangler $u$ and an isometry $w$, distributed in layers according to Fig. \ref{fig:symMERA}(i). In the presence of the symmetry (\ref{eq:symm}), we choose these tensors to be invariant under $\mathcal{G}$ \cite{symmetry}, 
\begin{eqnarray}
	(V_g \otimes V_g) ~u~ (V_g\otimes V_g)^{\dagger} &=& u, \nonumber\\
	(V_g \otimes V_g \otimes V_g) ~w~ (V_g)^{\dagger} &=& w
	\label{eq:symmetric}
\end{eqnarray}
where $V_g$ acting on different indices may actually denote different (in general, reducible) representations of $\mathcal{G}$. The layers of disentanglers and isometries define a real space RG transformation and a sequence of increasingly coarse-grained lattices $\{\L, \L', \L'', \cdots\}$. Under coarse-graining, a local operator $o$ transforms according to the scaling super-operator $\S$ of Fig. \ref{fig:CoarseGrain}(v) for $g=\mathbb{I}$,
\begin{equation}
o  \stackrel{\S}{\longrightarrow}  o' \stackrel{\S}{\longrightarrow} o''  ~\cdots~
\label{eq:S}
\end{equation}
The scaling operators $\phi_{\alpha}$ and scaling dimensions $\Delta_{\alpha}$ are obtained from the eigenvalue decomposition of the scaling superoperator $\S$ \cite{Transfer,MERACFT},
\begin{equation}
	\S(\phi_{\alpha}) = \lambda_{\alpha} \phi_{\alpha}, ~~~~~\Delta_{\alpha} \equiv -\log_3 \lambda_{\alpha}.	
	\label{eq:SS}
\end{equation}

\emph{Non-local scaling operators.}--- In this work we consider the coarse-graining of non-local operators $\to_g$ of the form
\begin{equation}
	\to_g = \tGamma_g \otimes o, ~~~ \tGamma_g \equiv  \underbrace{\cdots V_g \otimes V_g \otimes V_g}_{\infty}
	\label{eq:tog}
\end{equation}
where $\tGamma_g$ is a semi-infinite string made of copies of $V_g$ and $o$ is a local operator attached to the open end of $\tGamma_g$. Notice that, under coarse-graining, $\to_g$ is mapped into another non-local operator ${\to_g}'$ of the same type,
\begin{equation}
	\to_g = \tGamma_g \otimes o \longrightarrow {\to_g}' = \tGamma_g \otimes o',
\label{eq:Stog}
\end{equation}
since the semi-infinite string $\Gamma_{g}$ commutes with the coarse-graining everywhere except at its open end, as illustrated in Fig. \ref{fig:CoarseGrain}, where we exploit that the disentangler $u$ and isometry $w$ have been chosen to be symmetric, Eq. \ref{eq:symmetric}. In other words, we can study the sequence of coarse-grained non-local operators $\to_g \longrightarrow {\to_g}' \longrightarrow {\to_g}'' \cdots$ by just coarse-graining the operator $o$ with the modified scaling superoperator $\mathcal{S}_{g}$ of Fig. \ref{fig:CoarseGrain},
\begin{equation}
o  \stackrel{\S_{g}}{\longrightarrow}  o' \stackrel{\S_{g}}{\longrightarrow} o''  ~\cdots~
\label{eq:Stog3}
\end{equation}
In particular, by diagonalizing this scaling superoperator,
\begin{equation}
	\S_{g}(\phi_{g,\alpha}) = \lambda_{g,\alpha} \phi_{g,\alpha}~, ~~~~~\Delta_{g,\alpha} \equiv -\log_3 \lambda_{g,\alpha}~,	
\end{equation}
we obtain non-local scaling operators $\tphi_{g,\alpha}$ of the form
\begin{equation}
	\tphi_{g,\alpha} = \tGamma_g \otimes \phi_{g,\alpha}.
\end{equation}
Notice that for $g = \mathbb{I}$ we recover the local scaling operators $\phi_{\alpha}$ of Refs. \cite{Transfer,MERACFT}.

\emph{Quantum Ising model.}---As a first example, we use the above formalism to identify the non-local operator content of the Ising CFT starting from the Ising quantum spin chain, as described by the Hamiltonian
\begin{equation}
	\HI \equiv \sum_{r=-\infty}^{\infty} \left( X(r)X(r+1) + Z(r+1) \right),
\end{equation}
where $X$ and $Z$ are Pauli matrices. This model preserves parity, $\mathcal{G} = \mathbb{Z}_2$, 
so that $g\in \{+1,-1\}$, with $V_{+1} = \mathbb{I}$ and $V_{-1} = Z$, and
\begin{equation}
	\Gamma_{-1} ~\HI~ \Gamma_{-1}^{\dagger} = \HI,~~~~~\Gamma_{-1} \equiv \bigotimes_{m=-\infty}^{\infty} Z.
\end{equation}
Each index $i$ of tensors $u$ and $w$ decomposes as $i=(p,\alpha_p)$, where $p$ labels the parity ($p=1$ for even parity and $p=-1$ for odd parity) and $\alpha_p$ labels the distinct values of $i$ with parity $p$. Then the tensors $u$, $w$ are chosen to be parity preserving, e.g. $u_{i_1,i_2}^{j_1,j_2} = 0$ if $p(i_1)p(i_2)p(j_1)p(j_2)=-1$. 
An operator $O$ acting on the spin chain has parity $p$ if $(\Gamma_{-1}) ~ O ~ (\Gamma_{-1})^{\dagger} = p~ O$. 

\begin{figure}[!htbp]
\begin{center}
\includegraphics[width=8cm]{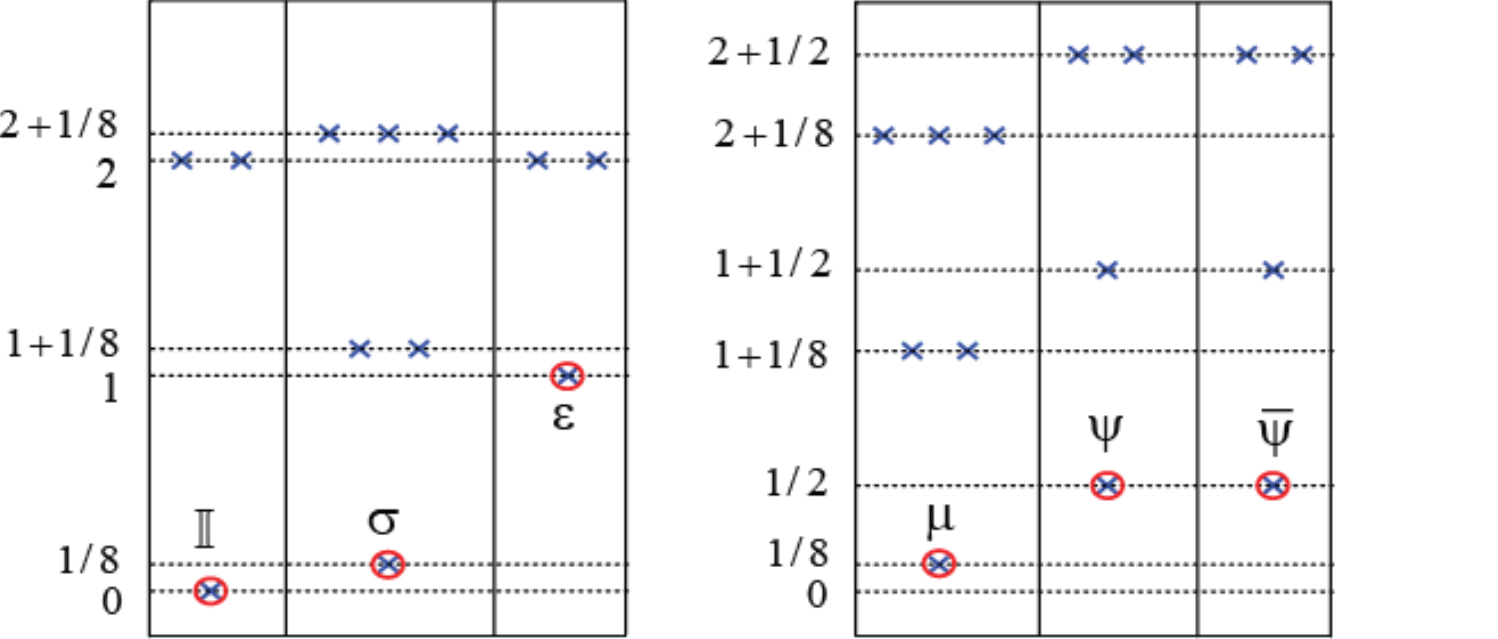}
\caption{ A few scaling dimensions of local (left) and non-local (right) scaling operators of the quantum Ising model, organized in its six conformal towers.}
\label{fig:IsingCritExp}
\end{center}
\end{figure}

We have used the algorithm of Refs. \cite{MERACFT,MERAalgorithm} to obtain a scale invariant MERA approximation for the ground state of $\HI$, with a computational effort that scales as $O(\chi^4\tilde{\chi}^4)$ with the dimension $\chi$ (and $\tilde{\chi}$) of the lower (and upper) indices of the disentangler $u$. The present results correspond to $\chi=36$ and $\tilde{\chi}=20$ and required one week on a 3 GHz dual core desktop with 8 Gb of RAM. The scaling superoperators $\S_{1}$ and $\S_{-1}$ were diagonalized in each parity sector. The resulting non-local scaling operators are of the form
\begin{equation}
	\tphi_{-1,\alpha} = \cdots Z\otimes Z \otimes Z \otimes \phi_{-1,\alpha}.
\end{equation}
Table \ref{tab:IsingExp} contains a few scaling dimensions extracted from $\S_{-1}$. The second and fifth columns are for scaling operators with even and odd parity, respectively, and reproduce the exact results with several digits of accuracy. Fig. \ref{fig:IsingCritExp} shows scaling dimensions for both local and non-local operators. Local scaling operators with even parity form the two conformal towers \cite{CFT} of the primary fields identity $\mathbb{I}$ and energy $\epsilon$ of the Ising CFT, whereas those with odd parity form the conformal tower of the primary field spin $\sigma$. Non-local scaling operators with even parity form the conformal tower of the disorder operator $\mu$, and those with odd parity are organized according to two towers corresponding to the fermion operators $\psi$ and $\bar{\psi}$.

\begin{table}[!htbp]
\begin{tabular}{||r|l|l||r|l|l||}
  \hline
  $\Delta^{\mbox{\tiny exact}}$  & $~\Delta^{\mbox{\tiny MERA}}_{\mbox{\tiny $\chi=36$}}~$ & error &
  $~\Delta^{\mbox{\tiny exact}}$ & $\Delta^{\mbox{\tiny MERA}}_{\mbox{\tiny $\chi=36$}}~$ & error\\ \hline
   $(\mu)$   1/8 &  $~$ 0.1250002     & 0.0002$\%$ & $(\psi)$ 1/2    &  $~$ 0.5      & $<10^{-8} \%$    \\

    1+1/8  &  $~$ 1.124937      & 0.006 $\%$ &    1+1/2  &   1.49999  & $<10^{-5} \%$    \\

    1+1/8  &  $~$ 1.124985      & 0.001 $\%$ &   2+1/2  &  2.49931  & 0.028 $\%$                \\

    2+1/8  &  $~$ 2.123237      & 0.083 $\%$ &    2+1/2  &  2.50118  & 0.047 $\%$                \\

    2+1/8  &  $~$ 2.124866      & 0.006 $\%$ &           &            &                           \\
   
   2+1/8  &  $~$ 2.125487      & 0.023 $\%$ &           &            &                           \\
  \hline
  \end{tabular} 
  \caption{Scaling dimensions of a few non-local operators of the quantum Ising model. The conformal towers of $\psi$ and $\bar{\psi}$ have identical scaling dimensions.}
  \label{tab:IsingExp}
\end{table}

We have also computed the coefficients $C_{\alpha\beta\gamma}$ of the operator product expansion (OPE) \cite{CFT} for all primary fields, by analysing three-point correlators as explained in Ref. \cite{MERACFT}. Notice that a three-point correlator $\langle \tphi_{g_1\alpha_1} \tphi_{g_2\beta} \tphi_{g_3\gamma} \rangle$ will vanish unless (i) the product of parities of the three operators is $+1$ (since the ground state is invariant under parity) and (ii) $g_1g_2g_3=\mathbb{I} \in \mathcal{G}$ (since otherwise the product $\tphi_{g_1\alpha_1} \tphi_{g_2\beta} \tphi_{g_3\gamma}$ is a non-local operator $\to$, which must decompose as a sum of non-local scaling operators $\tphi$, and $\langle \tphi \rangle = 0$ since all non-local scaling dimensions $\Delta_{-1,\alpha}$ are larger than zero, so that $\langle \to \rangle = 0$). Table \ref{tab:OPE} shows a numerical estimate of all non-vanishing OPE coefficients $C_{\alpha\beta\gamma}$. Again, the results match the exact solution with several digits of accuracy.

\begin{table}[!htbp]
\begin{tabular}{||l|l|l||}
  \hline
  $~~~~~C^{\mbox{\tiny exact}}$  & $~C^{\mbox{\tiny MERA}}_{\mbox{\tiny $\chi=36$}}~$ & error \\ \hline

   $C_{\epsilon, \sigma, \sigma} = 1/2$  &  $~$ 0.50008  & 0.016$\%$ \\
   $C_{\epsilon, \mu, \mu} = -1/2$  &  $~$ -0.49997  & 0.006$\%$    \\

   $C_{\psi,     \mu,    \sigma} = \frac{e^{-i\pi/4}}{\sqrt{2}}$     &  $\frac{1.00068e^{-i\pi/4}}{\sqrt{2}}$ $~$ & 0.068$\%$ \\ 
   $C_{\bar\psi,     \mu,    \sigma} = \frac{e^{i\pi/4}}{\sqrt{2}}$  &  $\frac{1.00068e^{i\pi/4}}{\sqrt{2}}$ $~$  & 0.068$\%$   \\

   $C_{\epsilon, \psi, \bar\psi} = i$     &  $1.0001 i$  $~$ & 0.010$\%$ \\
   $C_{\epsilon, \bar\psi, \psi} = -i$    &  $-1.0001 i$ $~$ & 0.010$\%$   \\

  \hline
  \end{tabular} \nonumber
  \caption{OPE coefficients for the local and non-local primary fields of the Ising CFT.}
  \label{tab:OPE}
\end{table}

Thus, not only have we been able to identify the \emph{entire} field content $\{\mathbb{I},\epsilon,\sigma,\psi,\bar{\psi},\mu\}$ of the Ising CFT from a simple and rather unexpensive analysis of a quantum spin chain, but we can now also identify all possible subsets of primary fields that close a subalgebra by inspecting Table \ref{tab:OPE}. Indeed, it follows that we have the following fusion rules
\begin{eqnarray}
 \epsilon \times \epsilon = \mathbb{I}, ~~~\sigma \times \sigma = \mathbb{I} + \epsilon,~~~ \sigma \times \epsilon = \sigma, \\
\mu \times \mu = \mathbb{I} + \epsilon, ~~~\mu \times \epsilon = \mu, \\
\psi \times \psi = \mathbb{I},~~~\bar{\psi} \times \bar{\psi} = \mathbb{I},\\
\psi \times \bar{\psi} = \epsilon,~~~
\psi \times \epsilon = \bar{\psi}, ~~~\bar{\psi} \times \epsilon = \psi,
\end{eqnarray}
(as well as other, such as $\sigma \times \mu = \psi + \bar{\psi}$, etc)
from where we see that $\{\mathbb{I},\epsilon\}$ and $\{\mathbb{I},\epsilon,\sigma\}$ close subalgebras of local primary fields, whereas $\{\mathbb{I},\epsilon,\mu\}$ and $\{\mathbb{I},\epsilon,\psi,\bar{\psi}\}$ close subalgebras that contain both local and non-local primary fields, where locality is relative to the spin variables.

\begin{figure}[!htbp]
\begin{center}
\includegraphics[width=7cm]{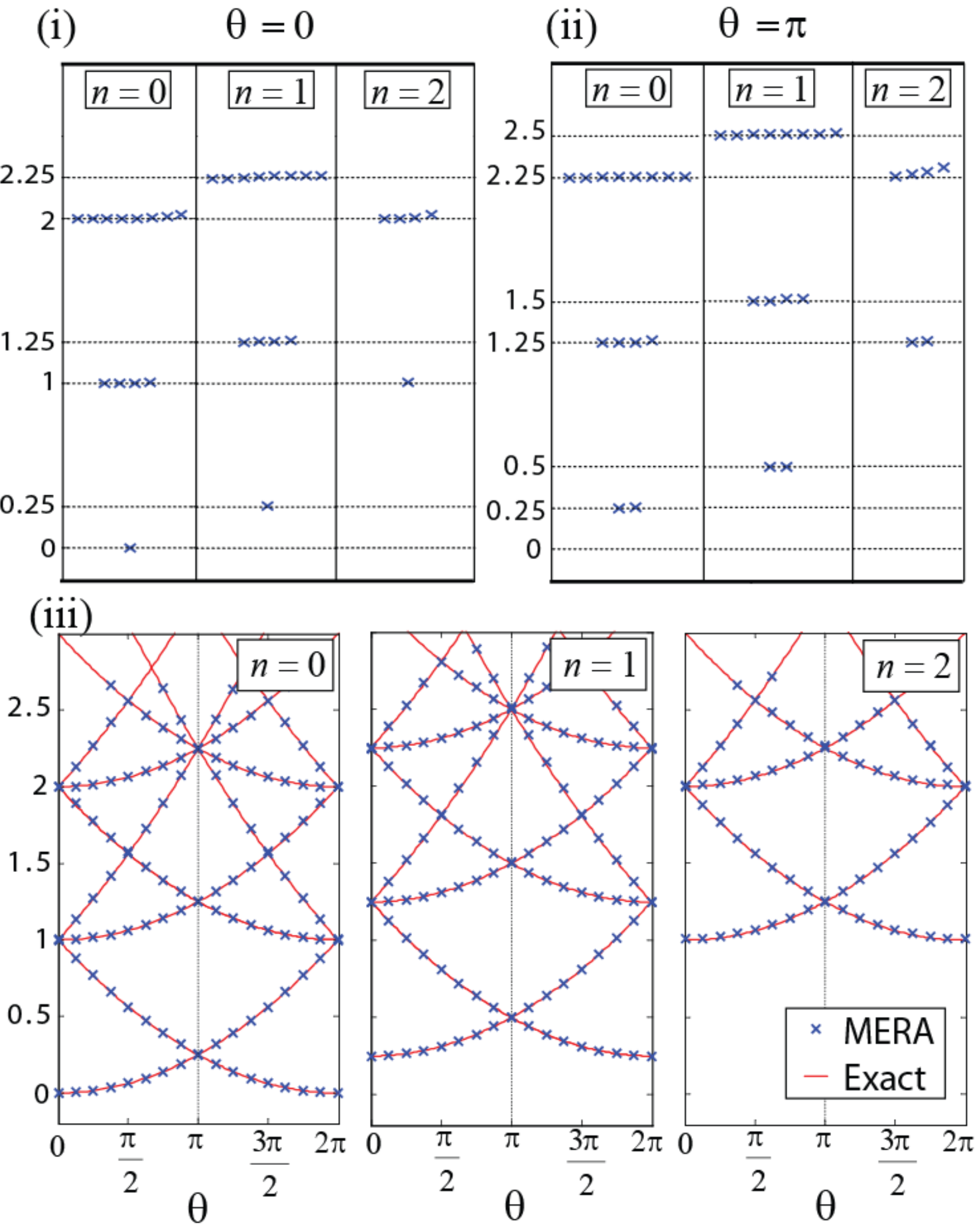}
\caption{ (i) Some scaling dimensions for \emph{local} operators of the quantum XX model. [Sectors with particle numbers $+|n|$ and $-|n|$ yield the same scaling dimensions.] (ii) Some scaling dimensions for \emph{non-local} operators with $V_{\theta} = Z$. (iii) Scaling dimensions $\Delta _{\theta,\alpha}$ as a function of $\theta$, see Eq. \ref{eq:exactTheta}.
The scaling dimensions for $n=0,1,2$ appear to only differ by a shift. }
\label{fig:XXCritExp}
\end{center}
\end{figure}

\emph{Quantum XX model}.---As a second example we study the quantum spin chain with Hamiltonian
\begin{equation}
	\HXX \equiv \sum_{r=-\infty}^{\infty} \left( X(r)X(r+1) + Y(r)Y(r+1)\right),
\end{equation}
where $X$ and $Y$ are Pauli matrices. This model is invariant under rotations $V_{\theta} = e^{-i\theta Z/2}$ on all spins. Therefore $\mathcal{G}=U(1)$, group elements $g$ can be labeled by an angle $\theta \in [0,2\pi)$, and
\begin{equation}
	\Gamma_{\theta} ~ \HXX ~ \Gamma_{\theta}^{\dagger} = \HXX,~~~~~\Gamma_{\theta} \equiv \bigotimes_{m=-\infty}^{\infty} V_{\theta}.
	\label{eq:exactTheta}
\end{equation}
To simplify the analysis, we regard each site as containing two spins, so that the on-site $\hat{z}$-component of the spin can take the values $0$ and $\pm 1$. Then each index $i$ of a tensor decomposes as $i=(n,\alpha_n)$, where $n\in \mathbb{Z}$ is the `particle number' (z spin component) and $\alpha_n$ labels the distinct values of $i$ with particle number $n$. Tensors $u$ and $w$ are chosen to be invariant under $U(1)$, e.g. $u^{j_1j_2}_{i_1i_2} = 0$ if $n(i_1)+n(i_2) \neq n(j_1)+n(j_2)$. An operator $O$ acting on the spin chain has particle number $n$ if $(\Gamma_{\theta}) O (\Gamma_{\theta})^{\dagger} = e^{-i n\theta}$.

The optimization of a scale invariant MERA with $\chi=54$ and $\tilde{\chi}=32$ took one week (by exploiting the block structure of the tensors \cite{symmetry}). For several values of $\theta \in [0,2\pi)$, we diagonalized the scaling superoperator $\S_{\theta}$ in the particle number sectors $n=0,\pm 1,\pm 2, \cdots$. Fig. \ref{fig:XXCritExp}(i)-(ii) show the resulting scaling dimensions $\Delta_{\theta, \alpha}$ for $\theta=0$ and $\pi$---that is for $V_{0}=\mathbb{I}$ (local operators) and $V_{\pi}=Z$, which also appear clearly organized in conformal towers. Finally, Fig. \ref{fig:XXCritExp}(iii) shows the scaling dimensions as a function of $\theta$. They are seen to accurately approximate the expression (denoted 'exact' in Fig. \ref{fig:XXCritExp}(iii))
\begin{equation} 
\Delta _{\theta,\alpha}  = \Delta _{0,\alpha}  + 
{\left( {\frac{\theta }{{2\pi }} + q} \right)^2  - q^2 } 
,\phantom{aaa} q=0,\pm 1,
\end{equation}
which is consistent with previous results \cite{XXTwist}. Up to a shift and a rescaling factor, the scaling dimensions $\Delta _{\theta,\alpha}$ reproduce the low energy spectrum of the XX chain with twisted boundary conditions with twisting angle $\theta$.

In summary, we have explained how to use the scale invariant MERA to characterize non-local scaling operators of a critical quantum spin chain. For the quantum Ising model, we have identified all non-local primary fields and obtained remarkably accurate estimates of their scaling dimensions and OPE coefficients. 
For the quantum XX model, we have obtained continuous families of non-local operators associated to twisted boundary conditions.

Support from the Australian Research Council (APA, FF0668731, DP0878830) is acknowledged.


\begin{thebibliography}{99}

\bibitem{ER} G. Vidal, Phys. Rev. Lett. \textbf{99}, 220405 (2007).
\bibitem{MERA} G. Vidal, Phys. Rev. Lett. \textbf{101}, 110501 (2008). 
\bibitem{Wilson} K.G. Wilson, Rev. Mod. Phys. \textbf{47}, 773 (1975).
\bibitem{DMRG} S. R. White, Phys. Rev. Lett. {\bf 69}, 2863 (1992).
 U. Schollwoeck, Rev. Mod. Phys. 77, 259 (2005)
\bibitem{Topo} M. Aguado, G. Vidal, Phys. Rev. Lett. \textbf{100}, 070404 (2008). 
R. Koenig, B. Reichardt, G. Vidal, Phys. Rev. B \textbf{79}, 195123 (2009).
\bibitem{Free} G. Evenbly, G. Vidal, Phys. Rev. B \textbf{81}, 235102 (2010);
ibid, New J. Phys. 12, 025007 (2010). 
\bibitem{Transfer} V. Giovannetti, S. Montangero, R. Fazio, Phys. Rev. Lett. \textbf{101}, 180503 (2008).
\bibitem{MERACFT} R. N. C. Pfeifer, G. Evenbly, G. Vidal, Phys. Rev. A \textbf{79}(4), 040301(R) (2009).
\bibitem{Fazio} S. Montangero et al., Phys. Rev. B \textbf{80}, 113103 (2009). 
V. Giovannetti et al., Phys. Rev. A \textbf{79}, 052314 (2009).
\bibitem{Boundary} G. Evenbly et al, arXiv:0912.1642.
\bibitem{symmetry} S. Singh, R. N. C. Pfeifer, G. Vidal, arXiv:0907.2994.
\bibitem{CFT} P. Di Francesco, P. Mathieu, and D. Senechal, \emph{Conformal Field Theory} (Springer, 1997). M. Henkel, \emph{Conformal Invariance and Critical Phenomena} (Springer, 1999).
\bibitem{MERAalgorithm} G. Evenbly, G. Vidal, Phys. Rev. B \textbf{79}, 144108 (2009). 
\bibitem{XXTwist} F.C. Alcaraz, M.N. Barber, M. T. Batchelor, Ann. Phys. \textbf{182}, 280 (1988). A. Kitazawa, J. Phys. A: Math. Gen. \textbf{30}, L285 (1997).

\end{thebibliography}
\end{document}